\newif\ifsubmode
\submodefalse

\ifsubmode
  \documentclass[12pt,preprint]{aastex}
  \received{}
  \revised{}
  \accepted{}
\else
  \documentclass[12pt,preprint]{emulateapj}

\fi

\newenvironment{inlinefigure}{
\def\@captype{figure}
\noindent\begin{minipage}{0.999\linewidth}\begin{center}}
{\end{center}\end{minipage}\smallskip}

\newcommand{\figonecap}{\label{fig1}
Size vs. redshift relation. Mean SExtractor half-light radii are
plotted with errorbars indicating the standard error of the mean
(i.e. the sample standard devation divided by the square-root of
the sample size).
The solid blue curve shows the expected trend
in the WMAP cosmology
if physical (proper) sizes do not evolve. The dashed red curve shows
the trend if sizes evolve as $H^{-1}(z)$ and the dotted green curve
shows $H^{-2/3}(z)$. The curves are all normalized to the mean
size at $z \approx 4$ (approximately $r_h = 1.7$ kpc).
}

\newcommand{\figtwocap}{
\label{fig2}                                                           
Observed size distributions. Data for the different samples are shown
with Poisson error bars. The curves show a ``no-evolution'' model
of the size distribution, tuned to match approximately the observed
distribution at $z \approx 4$. The model has a log-normal distribution
of radii with a peak at 2.1 kpc (for $L=2.9L^*$) and a width $\sigma =
0.3$. We adopt the size--luminosity relation ($R \propto L^{1/3}$),
observed for local disks \citep{2000ApJ...545..781D}.  The models
include a 50/50 mix of oblate ellipticals and flat exponential disks.
Model galaxies have been inserted into the images in a Monte-Carlo
fashion and detected and measured using SExtractor in the same way
as the real galaxies. Therefore the model curves faithfully represent
the effects of both sample incompleteness and measurement biases. The
input distribution (prior to selection and measurement biases) for the
$z \approx 4$ bin is shown as the dashed curve.  Objects classified by
SExtractor neural-network as stellar at greater than 90\% probability
are excluded from the samples (both the data and the simulations). In
practice only a few objects with $r_h < 0.1$\arcsec\ are removed.
}

\newcommand{\figthreecap}{
\label{fig3}
(a) Ellipticity distribution for $z \approx 4$ galaxies compared to
simulations for pure disk and pure spheroid populations (the blue-dashed
and red-solid curves, respectively).  For the simulations, the disk sample
(viewed from arbitrary directions) is drawn from a population of oblate
optically-thin spheroids with a Gaussian distribution of intrinsic axial
ratios with mean $b/a = 0.05$ and $\sigma = 0.01$.  The spheroids are
drawn from a population of oblate spheroids with intrinsic axial ratios
uniformly populating the range $0.3 < b/a < 0.9$.  The samples for both
observations and simulations are limited to galaxies with $23 < I_{775}
< 25$ and SExtractor half-light radii $r_{h} > 0.2$\arcsec.

(b) Concentration-index distribution for
for $z \approx 4$ galaxies compared to simulations for
pure disk and pure spheroid populations (the blue-dashed and
red-solid curves, respectively). Samples are the same
as in (a).
}

\shorttitle{High-Redshift Galaxy Sizes}
\shortauthors{Ferguson et al.}

\begin{document}

\title{The Size Evolution of High Redshift Galaxies\altaffilmark{1}}

\author{
Henry C. Ferguson\altaffilmark{2,3}, 
Mark Dickinson\altaffilmark{2,3}, 
Mauro Giavalisco\altaffilmark{2,11}, 
Claudia Kretchmer\altaffilmark{3}, 
Swara Ravindranath\altaffilmark{2}, 
Rafal Idzi\altaffilmark{3}, 
Edward Taylor\altaffilmark{4}, 
Christopher J. Conselice\altaffilmark{5}, 
S. Michael Fall\altaffilmark{2}, 
Jonathan P. Gardner\altaffilmark{6}, 
Mario Livio\altaffilmark{2}, 
Piero Madau\altaffilmark{7}, 
Leonidas A. Moustakas\altaffilmark{2}, 
Casey M. Papovich\altaffilmark{8}, 
Rachel S. Somerville\altaffilmark{2}, 
Hyron Spinrad\altaffilmark{9}, 
Daniel Stern\altaffilmark{10,11}
}

\altaffiltext{1}{Based on observations obtained with the NASA/ESA
{\it Hubble Space Telescope} (HST), the European Southern Observatory,  
and the Kitt Peak National Observatory (KPNO). HST is operated by the
Association of Universities for Research in Astronomy, Inc. (AURA)
under NASA contract NAS5-26555. KPNO is part of the National Optical
Astronomy Observatories, which is operated also by AURA under cooperative
agreement with the National Science Foundation.}

\altaffiltext{2}{Space Telescope Science Institute, 3700 San Martin Drive, 
Baltimore, MD  21218, USA}

\altaffiltext{3}{Department of Physics and Astronomy, The Johns Hopkins
University, 3400 N. Charles St., Baltimore, MD 21218}

\altaffiltext{4}{University of Melbourne, Australia}

\altaffiltext{5}{Caltech, MS 105-24, Pasadena CA}

\altaffiltext{6}{Laboratory for Astronomy and Solar Physics, Code 681, 
Goddard Space Flight Center, Greenbelt MD 20771}

\altaffiltext{7}{Department of Astronomy and Astrophysics, University
of California, Santa Cruz, CA 95064.}

\altaffiltext{8}{University of Arizona}

\altaffiltext{9}{
Dept. of Astronomy, University of California, Berkeley Berkeley CA. 94720-3411}

\altaffiltext{10}{Jet Propulsion Laboratory, California Institute of Technology,
Mail Stop 169-506, Pasadena, CA  91109, USA }

\altaffiltext{11}{Visiting astronomer Kitt Peak National Observatory.}

\begin{abstract}
Hubble Space Telescope images of high-redshift galaxies selected via color
and photometric redshifts are used to examine the size and axial-ratio
distribution of galaxies as a function of redshift at lookback times $t
> 8$ Gyr. These parameters are measured at rest-frame UV wavelengths
($1200< \lambda <2000$ \AA) on images with a rest-frame resolution
of less than 0.8 kpc.  Galaxy radii are found to scale with redshift
approximately as Hubble parameter $H^{-1}(z)$. This is in accord with
the theoretical expectation that the typical sizes of the luminous
parts of galaxies should track the expected evolution in the virial
radius of dark-matter halos.  The mean ratio of semi-major to
semi-minor axis for a bright well-resolved sample of galaxies
at $z \sim 4$ is $b/a =0.65$, suggesting that these Lyman 
break galaxies are not drawn from a spheroidal population. 
However the median concentration index of this sample is $C = 3.5$,
which is closer to the typical concentration indices, $C \sim 4$, of nearby
elliptical galaxies than to the values, $C<2 $ for local disk galaxies
of type Sb and later.

\end{abstract}


\section {Introduction}

An important goal of cosmology is to understand how galaxies evolve
toward their current sizes and shapes.  The basic framework of galaxy
formation within the hierarchical Cold-Dark-Matter (CDM) cosmology was
set out by \citet{WR78}, and has been refined by numerous N-body and
semi-analytical studies \citep{WF91, KWG93, CAFNZ94, 1999MNRAS.310.1087S}.
The formation of galactic disks within dark-matter halos was studied by
\citet{1980MNRAS.193..189F}. In this model, dark halos acquire their
angular momenta via tidal torques, the angular momentum per unit mass
of the baryons and the dark matter are initially the same, and angular
momentum is conserved as the baryons collapse and cool to form a disk.
With these simplifying assumptions the baryons typically collapse by
factors of $\sim 10$ and the resulting disks have rotation curves,
surface-density profiles, and scale radii similar to those observed.
Further analytical studies have calculated the distribution function of
disk-galaxy sizes and the disk-galaxy size-redshift relation \citep{DSS97,
MMW98}. Meanwhile cosmological N-body + hydrodynamical simulations have
formed disks that appear similar to spiral galaxies, but that tend to
be too small to match present-day galaxies (\citealt{NavSteinmetz97};
but see \citealt{2000MNRAS.315L..18E}).  It is not yet known whether
the source of this discrepancy lies in baryonic, dark-matter physics,
or numerical issues in the simulations.

If the general Fall \& Efstathiou view of disk formation is correct,
there are several rather robust expectations that are worth exploring
through the observations of high-redshift galaxies.

{\em The size-redshift relation.}
The sizes of galactic disks forming at a redshift $z$ should be
a fixed fraction of the size of the dark-matter halo. The virial-radius
of a dark-matter halo scales with redshift and virial velocity 
$V_{\rm vir}$ or virial mass $M_{\rm vir}$ as 
\begin{equation}
R_{\rm vir} = \left[ {{GM_{\rm vir}} \over {100 H^2(z)}} \right]^{1/3} =
{{V_{\rm vir}} \over {10 H(z)}},
\end{equation}
where
\begin{equation}
H(z) = H_0 [ \Omega_m (1+z)^3 + \Omega_k (1+z)^2 + \Omega_\Lambda ] ^{1/2}
\end{equation}
is the Hubble parameter at redshift $z$ \citep{CPT92}.
Assuming that the exponential scale-length $R_s$ of the baryonic disk
scales with the virial radius, the sizes of disks are expected to
scale with redshift as $R_s \propto H^{-1}(z)$ at fixed circular
velocity, or $R_s \propto H^{-2/3}(z)$ at fixed mass. Observations
are more suitable to tracking evolution with fixed luminosity $L$, which
may be expected to fall somewhere in between these two functional forms.
Previous observations have demonstrated that high-redshift galaxies are
small (e.g. \citealt{LickHDFhighz}; \citealt{BBI2003}). The
larger sample of galaxies provided by GOODS allows a closer look
at the trend with redshift.

{\em The size distribution.}
If governed primarily by the angular momenta of their halos, the sizes
of disk galaxies should 
should show a log-normal distribution proportional
to the dimensionless spin parameter:
\begin{equation}
p(\lambda) d \lambda = {{1} \over {\sqrt{2 \pi \sigma_\lambda}}}
\exp \left[ - {{\ln^2 (\lambda/\bar{\lambda})} \over {2 \sigma_\lambda^2}} \right]
{{d \lambda} \over {\lambda}}, 
\end{equation}
where $\lambda$ is related to the total angular momentum $J$, energy $E$,
and mass $M$ of the dark matter halo through the definition
$\lambda \equiv J |E|^{1/2} G^{-1} M^{-5/2}$. 
The predicted functional form for the distribution of disk-galaxy sizes 
appears to be a reasonable match to observations of present-day
disk galaxies, albeit with
a width $\sigma$ that is significantly smaller than
predicted by tidal-torque theory \citep{2000ApJ...545..781D}. 
It is sensible to imagine that a simple proportionality between galaxy
radius and halo radius might hold even for elliptical galaxies, although
models are less well developed in this case. It is less 
obvious that the size distribution should show the same functional form.

{\em The ellipticity distribution.}
The theory described above is specific to disk galaxies. It is thus
interesting to test whether or not high-redshift galaxies are predominantly
disks. A relatively straightforward test is to compare the observed distribution
of axial ratios to the distribution expected for disks or spheroids 
viewed at a random set of orientations, applying the classic test
outlined by \citet{1970ApJ...160..831S}.  

Throughout this paper we use a cosmology with parameters 
$h,\Omega_{\rm tot},\Omega_{m},\Omega_{\Lambda} = 0.7,1.0,0.3,0.7$,
in the notation of \citet{CPT92}. The sizes of galaxies quoted
in this paper are in physical (proper), not co-moving, units.
All magnitudes are on the AB system \citep{Oke74}.

\section{Observations and Measurements}\label{observations}

The Chandra Deep Field South (CDF-S) and Hubble Deep Field North (HDF-N) 
were observed with the Hubble Space Telescope (HST) Advanced Camera
for Surveys (ACS) as part of the Great Observatories Origins Deep Survey
(GOODS).  For the HDF-N, U-band observations were carried
out using the Mosaic camera on the 4-m Mayall telescope.
Details of the observations, data reduction, and catalog generation
are described in \citet{Giavalisco2003a}. Samples of galaxies in
broad bins of redshift were constructed using
the \citet{Giavalisco2003b} Lyman-break selection for $z>3$
and the \citet{Mobasher2003b} CDF-S photometric redshifts 
(for two bins spanning the range $z=1 - 1.9$ and $z=1.9-2.8$).
The $z \approx 3$ and $4$ samples use data from both the HDF-N and the CDF-S. The 
$z \approx 3$ sample is drawn only from the HDF-N, where deep U-band images exist.
Because the redshift selection is based entirely on photometry, there
is undoubtedly some scatter of galaxies from one redshift bin to
another, and a few interlopers from lower redshifts in the high redshift bins.
Based on simulations of template galaxies of various spectral types
observed at the GOODS S/N, we expect the contamination 
to affect only the tails of the observed
size, ellipticity, and concentration-index distributions. Spectroscopic
confirmations are of course highly desirable.

Two different techniques were used to measure the radii of the galaxies.
Half-light radii (denoted by $r_h$) were measured using SExtractor \citep{BA96}. 
For this analysis SExtractor performs circular-aperture photometry using
an analysis aperture radius that is ten times larger than 
the first radial-moment of the light distribution defined by connected
pixels more than 0.6 $\sigma$ brighter than the background in a smoothed
detection image.
\citet{1976ApJ...209L...1P} radii were measured using the prescription adopted for the
SDSS project \citep{2002AJ....123..485S}. 
The ratio of the semi-minor to semi-major axis $b/a$ is measured
by SExtractor using second moments of the light distribution within
the analysis aperture.


To compare the sizes of galaxies in different redshift intervals, we
confine our samples to non-evolving rest-frame luminosities between
$0.7L^*$ and $5 L*$, where $L^*$ is the characteristic luminosity
of a $z=3$ Lyman break galaxy from \citet{SAGDP99}.  Five samples
in different bins of redshift ($z \approx 1.4, 2.3, 3.0, 4.0,$ and
$5.0$) are constructed.  For each sample we select galaxies within the
desired luminosity range based on their apparent $z$-band magnitudes,
after accounting for cosmological effects and $k$-corrections assuming
a typical Lyman-break galaxy spectral-energy distribution (SED). The
adopted SED is from the 1999 version of the \citet{BC93} solar-metallicity models,
with a Salpeter initial-mass function, constant
star-formation rate, and age of 144 Myr.  This is attenuated assuming $E(B-V) =
0.14$ with the \citet{Calzetti2000} extinction curve to match the mean
UV spectral slope seen in the $z \sim 3$ samples \citep{2000ApJ...544..218A}.
The luminosity range translates to a $z$-band magnitude
range of 23.1 to 25.3 for the $z \sim 4$ sample, with that magnitude range
shifting by -2.41, -1.12, -0.64, and +0.46 magnitudes for the samples
in the redshift bins at $z \sim 1.4, 2.3, 3.0,$ and $5.0$, respectively.
After rejecting sources 
classified as stellar (SExtractor \verb!CLASS_STAR!$ > 0.9$) 
or with
unreliable photometry (SExtractor \verb!FLAG!$ > 3$), the sample sizes (for the
lowest to highest redshift bins, respectively) are 21, 73, 140, 386,
and 153 galaxies.

\section{Results}\label{results} 

The galaxy sizes are measured in the band closest to rest-frame
$1500${\AA}.  At $z = 3,4,$ and $5$, the central
wavelengths of the F606W, F775W, and F850LP band filters fall at
rest-frame $\lambda = 1500 \pm 40$ {\AA}. For the lower-redshift
samples the size measurement was made in the F435W images corresponding
to $\lambda \sim 1800$\AA\  at $z = 1.4$ and $\lambda \sim 1300$\AA\ at $z=2.3$.
The sizes for any of these samples do not change appreciably
when measured through the next redder bandpasses.

The measured size-redshift relation for our samples is shown in
Fig. \ref{fig1}. The observed points are uncorrected for biases
or incompleteness (see \S\ref{biases}), and the errorbars represent
the standard error of the mean for each sample, ignoring possible
effects of clustering \citep{Somerville2003a}
or crowding. The measurements are roughly
consistent with constant angular sizes above $z = 2.5$.
For comparison, in Fig. \ref{fig1} we show three size-redshift
relations. The blue line is the angular-size
redshift relation for a standard measuring-rod in the WMAP cosmology
\citep{Spergel2003}. In this cosmology, angular diameters increase
by 23\% from $z=3$ to $z=5$ (for objects with fixed proper diameters).
The two lines that decrease with redshift
are the expected scalings if size at fixed luminosity tracks 
size at fixed virial velocity $r \propto H^{-1}(z)$ or fixed mass
$r \propto H^{-2/3}(z)$. The mean half-light radius of 
0.24\arcsec\ at $z \approx 4$ corresponds to 1.7 kpc. 

\ifsubmode
\else
\begin{inlinefigure}
\begin{center}
\resizebox{\textwidth}{!}{\includegraphics{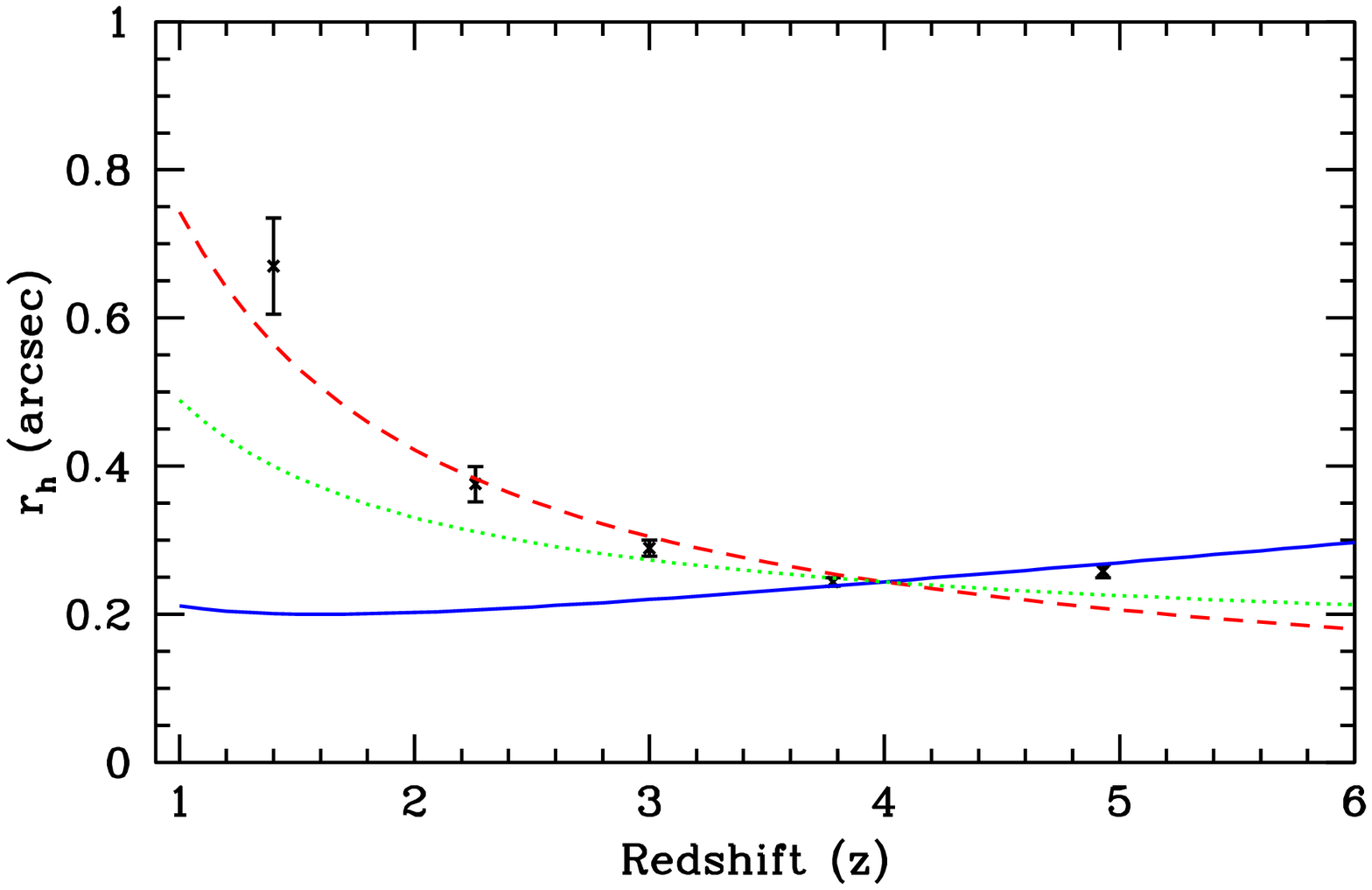}}
\figcaption{ \figonecap }
\end{center}
\end{inlinefigure}
\fi

The size distributions of galaxies in the different redshift bins
are shown in Fig. \ref{fig2}. The curves show a ``no-evolution''
model tuned to match the observed size distribution at $z=4$.
This model accounts for selection and measurement biases 
(including cosmological surface-brightness dimming) as
described in \S\ref{biases}. The galaxies in
the two low-redshift bins are clearly systematically larger than 
the no-evolution model would predict. However, given the sample
sizes and measurement biases, the trend between redshifts $z=3$ and $z=5$
is not easily distinguished from the no-evolution model.

Figure \ref{fig3}a compares the observed distribution of ellipticities
for the B-dropout $z \approx 4$ sample to simulations. 
The observed population has 
a larger proportion of galaxies with axial ratios $b/a \lesssim 0.4$
than expected for a purely spheroidal population. A Kolmogorov-Smirnov
test indicates a likelihood of 57\% that observed distribution could
have been drawn from the model disk distribution, while the probability
is $5\times10^{-7}$ that the observed distribution could have been
drawn from the model spheroidal distribution. Similar results
(at lower significance) are obtained for the $z \sim 3$ and $z \sim 5$ 
samples.  However, these statistical probabilities are sensitive to the 
size and magnitude distribution of galaxies fed into the simulations.
While we have attempted to match the observed distribution of
magnitudes and sizes (and colors) at $z=4$ (see Fig. 2), more work
is needed to verify that the preference for flattened systems is 
indeed robust.  Visual inspection of the images reveals
very few galaxies that have morphologies reminiscent of nearby 
disk galaxies. The observed flattening distribution is probably 
a reflection of the tendency for many of the  Lyman-break galaxies to have a few
concentrations of light of nearly equal brightness rather than a 
central dominant concentration.

\ifsubmode
\else
\begin{inlinefigure}
\begin{center}
\resizebox{\textwidth}{!}{\includegraphics{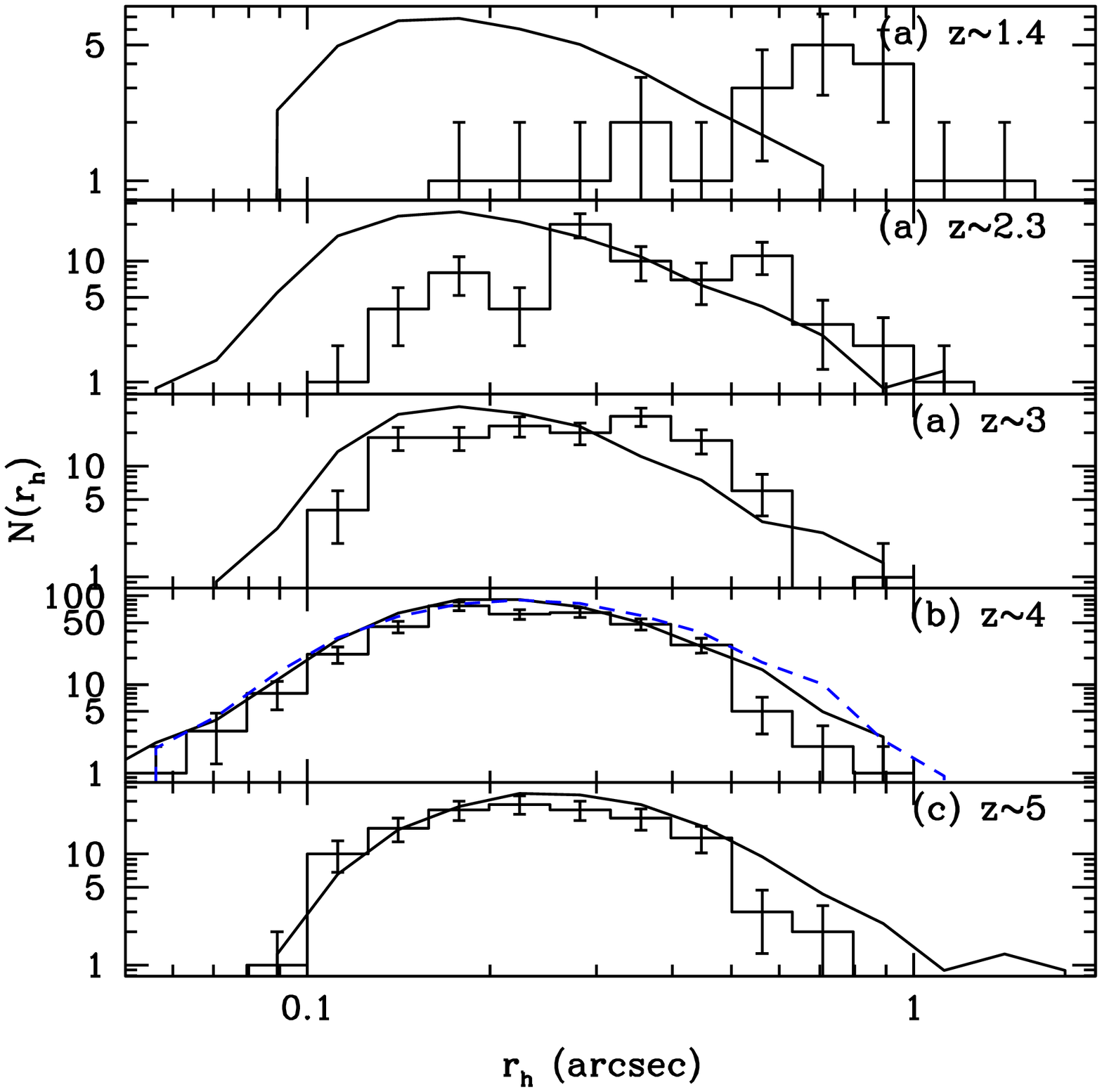}}
\figcaption{ \figtwocap }
\end{center}
\end{inlinefigure}
\fi

The concentration-index distribution for the same sample
is shown in Fig. \ref{fig3}b.  The concentration index here is defined
as $5 \log(r_{80}/r_{20})$ where $r_{80}$ and $r_{20}$ are radii
containing 80\% and 20\% of the galaxy flux, respectively
\citep{Conselice2003a}.
The observed galaxies span a broader range
than the ideal disks and spheroids that populate the simulations.
The $z \approx 4$ population tends to be more centrally concentrated than 
pure exponentials.
Nearby spiral galaxies tend to have central concentrations of light
as well, but these also tend to be red, and virtually invisible
in the rest-frame ultraviolet where the current measurements are made.

\ifsubmode
\else
\begin{inlinefigure}
\begin{center}
\resizebox{\textwidth}{!}{\includegraphics{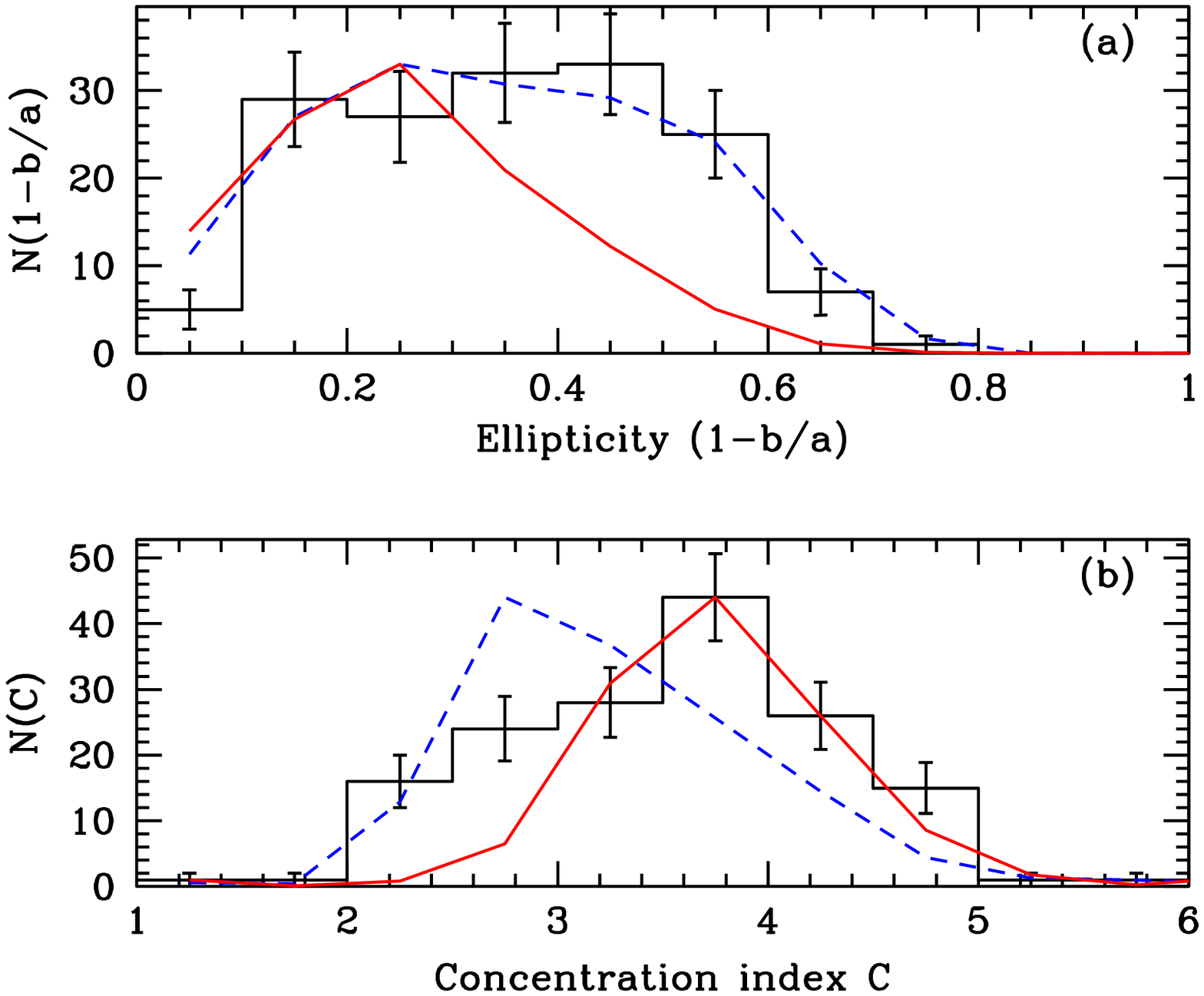}}
\figcaption{\figthreecap}
\end{center}
\end{inlinefigure}
\fi

\section{Selection Effects and Biases}\label{biases} 

In this section we highlight the issues of bias and incompleteness
and the tests that have
been carried out to verify the results of the previous sections.

{\it Incompleteness.} The completeness limits of the GOODS images
have been estimated by inserting artificial galaxies with a
wide range of half-light radii and magnitudes into the GOODS $z$-band
image, and re-running SExtractor. The simulated galaxies have either $r^{1/4}$-law
or exponential surface-brightness profiles, with the axial ratio
distributions described in the caption to Fig. 3.
The completeness limits as a function of
size and magnitude are shown by \citet{Giavalisco2003a}.
At $z_{850} = 24.7$, corresponding to the mean magnitude of
the $z \approx 4$ sample considered in this letter, the sample appears
to be 80\% complete for galaxies with half-light radii $r_h=0.5$\arcsec\
and 50\% complete for galaxies with $r_h = 0.9$\arcsec. 
The $z>3$ samples are affected by incompleteness for galaxies with sizes 
$r_h \gtrsim 3.5$ kpc; for the lower-redshift samples there is no
equivalent bias against detecting galaxies with small sizes. Therefore
the most robust result is the observation
that analogs of the galaxies with small half-light radii in the $z>3$ bins
are much less common at the lower redshifts. Without
careful modeling (or deeper images) it is not possible to rule out
the possibility that galaxies with larger sizes exist at the higher
redshifts. It is worth noting that a pure exponential disk galaxy with 
a luminosity $L^*$ and a scale length of $>5$ kpc would 
be unlikely to be detected at $z>3$ in the GOODS $z$-band images.

{\it Radius-measurement biases.} The results shown in Fig. 1 and 2 
are for SExtractor half-light radii. The trends and distributions
look virtually identical when Petrosian radii are used. Based
on the simulations it is clear that both radius measurements become
biased at faint magnitudes and small sizes. For example, for galaxies in the
magnitude range of our $z \approx 4$ sample, a galaxy with an
intrinsic $r_h = 0.5${\arcsec} would have a measured $r_h = 0.36$\arcsec.
Measurement biases will
thus pull in the tail of the size distribution; the true size
distributions at $z>3$ are almost certainly broader and peaked at slightly
larger sizes than the observed distributions shown in Fig. 2. 
The model distribution for the $z \sim 4$ sample provides an estimate
of the importance of this bias. The true distribution of half-light
radii input to the simulation is shown as the dashed line.
For this input distribution the effect of selection
and the size-measurement bias are minor. A distribution
with a more significant tail to large sizes would show more of an effect.
We have tried broader distributions and 
distributions with a larger mean size and found them to predict more
large galaxies than observed in the $z \approx 4$ sample.

\section{Summary and Conclusions}\label{conclusions} 

Samples of galaxies within a fixed and fairly
narrow range of rest-frame UV luminosities have been compared from
redshifts $z \sim 1 $ to $z \sim 5$. 
The sizes of high-redshift galaxies appear to evolve roughly
as $H^{-1}(z)$ in agreement with the rather robust expectation from
hierarchical theory. This general conclusion is subject to the
caveat that we may be seeing different kinds of galaxies at the
different redshifts. Our sample selection necessarily targets
UV-bright objects; further studies on samples selected
in the near-infrared or in subsets of morphology are clearly
warranted. For the galaxy sample discussed here, the size 
distribution at $z \sim 4$ is 
reasonably well represented by a log-normal
distribution, but with a smaller mean size than 
observed for samples of nearby luminous galaxies.
Measurements of the ellipticity and concentration-index
distributions suggest that Lyman-break galaxies at $z \sim 4$ represent a 
mix of morphologies, with some tendency toward flattened systems.

\acknowledgments

Support for this work was
provided by NASA through grant GO09583.01-96A from the Space Telescope
Science Institute, which is operated by the Association of
Universities for Research in Astronomy, under NASA contract
NAS5-26555.
Support for this work, part of the {\it Space Infrared Telescope 
Facility (SIRTF)} Legacy Science Program, was provided by NASA through 
Contract Number 1224666 issued by the Jet Propulsion Laboratory, 
California Institute of Technology under NASA contract 1407. 
PM acknowledges support by NASA through grant NAG5-11513.

\bibliographystyle{apj}
\bibliography{apjmnemonic,bib}

\begin{thebibliography}{}

\bibitem[\protect\citeauthoryear{{Adelberger} \& {Steidel}}{{Adelberger} \&
  {Steidel}}{2000}]{2000ApJ...544..218A}
{Adelberger}, K.~L.,  \& {Steidel}, C.~C. 2000, \apj, 544, 218

\bibitem[\protect\citeauthoryear{{Bertin} \& {Arnouts}}{{Bertin} \&
  {Arnouts}}{1996}]{BA96}
{Bertin}, E.,  \& {Arnouts}, S. 1996, A\&AS, 117, 393

\bibitem[\protect\citeauthoryear{{Bouwens}, {Broadhurst}, \&
  {Illingworth}}{{Bouwens} et~al.}{2003}]{BBI2003}
{Bouwens}, R.~J., {Broadhurst}, T.,  \& {Illingworth}, G. 2003,
  astro-ph/0304547

\bibitem[\protect\citeauthoryear{{Bruzual} \& {Charlot}}{{Bruzual} \&
  {Charlot}}{1993}]{BC93}
{Bruzual}, A.~G.,  \& {Charlot}, S. 1993, ApJ, 405, 538

\bibitem[\protect\citeauthoryear{{Calzetti} et~al.}{{Calzetti}
  et~al.}{2000}]{Calzetti2000}
{Calzetti}, D., {Armus}, L., {Bohlin}, R.~C., {Kinney}, A.~L., {Koornneef}, J.,
   \& {Storchi-Bergmann}, T. 2000, \apj, 533, 682

\bibitem[\protect\citeauthoryear{{Carroll}, {Press}, \& {Turner}}{{Carroll}
  et~al.}{1992}]{CPT92}
{Carroll}, S.~M., {Press}, W.~H.,  \& {Turner}, E.~L. 1992, ARA\&A, 30, 499

\bibitem[\protect\citeauthoryear{{Cole} et~al.}{{Cole} et~al.}{1994}]{CAFNZ94}
{Cole}, S., {Aragon-Salamanca}, A., {Frenk}, C.~S., {Navarro}, J.~F.,  \&
  {Zepf}, S.~E. 1994, MNRAS, 271, 781

\bibitem[\protect\citeauthoryear{{Conselice}}{{Conselice}}{2003}]{Conselice200%
3a}
{Conselice}, C. 2003, astro-ph/0303065

\bibitem[\protect\citeauthoryear{{Dalcanton}, {Spergel}, \&
  {Summers}}{{Dalcanton} et~al.}{1997}]{DSS97}
{Dalcanton}, J.~J., {Spergel}, D.~N.,  \& {Summers}, F.~J. 1997, ApJ, 482, 659

\bibitem[\protect\citeauthoryear{{de Jong} \& {Lacey}}{{de Jong} \&
  {Lacey}}{2000}]{2000ApJ...545..781D}
{de Jong}, R.~S.,  \& {Lacey}, C. 2000, \apj, 545, 781

\bibitem[\protect\citeauthoryear{{Eke}, {Efstathiou}, \& {Wright}}{{Eke}
  et~al.}{2000}]{2000MNRAS.315L..18E}
{Eke}, V., {Efstathiou}, G.,  \& {Wright}, L. 2000, \mnras, 315, L18

\bibitem[\protect\citeauthoryear{{Fall} \& {Efstathiou}}{{Fall} \&
  {Efstathiou}}{1980}]{1980MNRAS.193..189F}
{Fall}, S.~M.,  \& {Efstathiou}, G. 1980, \mnras, 193, 189

\bibitem[\protect\citeauthoryear{{Giavalisco} et~al.}{{Giavalisco}
  et~al.}{2003a}]{Giavalisco2003a}
{Giavalisco}, M.,  et~al. 2003a, this ApJL issue

\bibitem[\protect\citeauthoryear{{Giavalisco} et~al.}{{Giavalisco}
  et~al.}{2003b}]{Giavalisco2003b}
{Giavalisco}, M.,  et~al. 2003b, this ApJL issue

\bibitem[\protect\citeauthoryear{{Kauffmann}, {White}, \&
  {Guiderdoni}}{{Kauffmann} et~al.}{1993}]{KWG93}
{Kauffmann}, G., {White}, S. D.~M.,  \& {Guiderdoni}, B. 1993, MNRAS, 264, 201

\bibitem[\protect\citeauthoryear{{Lowenthal} et~al.}{{Lowenthal}
  et~al.}{1997}]{LickHDFhighz}
{Lowenthal}, J.~D., et~al. 1997, ApJ, 481, 673

\bibitem[\protect\citeauthoryear{{Mo}, {Mao}, \& {White}}{{Mo}
  et~al.}{1998}]{MMW98}
{Mo}, H.~J., {Mao}, S.,  \& {White}, S. D.~M. 1998, MNRAS, 295, 319

\bibitem[\protect\citeauthoryear{{Mobasher} et~al.}{{Mobasher}
  et~al.}{2003}]{Mobasher2003b}
{Mobasher}, B.,  et~al. 2003, this ApJL issue

\bibitem[\protect\citeauthoryear{{Navarro} \& {Steinmetz}}{{Navarro} \&
  {Steinmetz}}{1997}]{NavSteinmetz97}
{Navarro}, J.~F.,  \& {Steinmetz}, M. 1997, ApJ, 478, 13

\bibitem[\protect\citeauthoryear{{Oke}}{{Oke}}{1974}]{Oke74}
{Oke}, J.~B. 1974, ApJS, 27, 21

\bibitem[\protect\citeauthoryear{{Petrosian}}{{Petrosian}}{1976}]{1976ApJ...20%
9L...1P}
{Petrosian}, V. 1976, \apjl, 209, L1

\bibitem[\protect\citeauthoryear{{Sandage}, {Freeman}, \& {Stokes}}{{Sandage}
  et~al.}{1970}]{1970ApJ...160..831S}
{Sandage}, A., {Freeman}, K.~C.,  \& {Stokes}, N.~R. 1970, \apj, 160, 831

\bibitem[\protect\citeauthoryear{{Somerville} et~al.}{{Somerville}
  et~al.}{2003}]{Somerville2003a}
{Somerville}, R.~S.,  et~al. 2003, this ApJL issue

\bibitem[\protect\citeauthoryear{{Somerville} \& {Primack}}{{Somerville} \&
  {Primack}}{1999}]{1999MNRAS.310.1087S}
{Somerville}, R.~S.,  \& {Primack}, J.~R. 1999, \mnras, 310, 1087

\bibitem[\protect\citeauthoryear{{Spergel} et~al.}{{Spergel}
  et~al.}{2003}]{Spergel2003}
{Spergel}, D.,  et~al. 2003, astro-ph/0302209

\bibitem[\protect\citeauthoryear{{Steidel} et~al.}{{Steidel}
  et~al.}{1999}]{SAGDP99}
{Steidel}, C.~C., {Adelberger}, K.~L., {Giavalisco}, M., {Dickinson}, M.,  \&
  {Pettini}, M. 1999, ApJ, 519, 1

\bibitem[\protect\citeauthoryear{{Stoughton} et~al.}{{Stoughton}
  et~al.}{2002}]{2002AJ....123..485S}
{Stoughton}, C., et~al. 2002, \aj, 123, 485

\bibitem[\protect\citeauthoryear{{White} \& {Frenk}}{{White} \&
  {Frenk}}{1991}]{WF91}
{White}, S. D.~M.,  \& {Frenk}, C.~S. 1991, ApJ, 379, 52

\bibitem[\protect\citeauthoryear{{White} \& {Rees}}{{White} \&
  {Rees}}{1978}]{WR78}
{White}, S. D.~M.,  \& {Rees}, M.~J. 1978, MNRAS, 183, 341

\end{thebibliography}

\ifsubmode
\newpage

\begin{inlinefigure}
\begin{center}
\resizebox{\textwidth}{!}{\includegraphics{f1.ps}}
\figcaption{ \figonecap }
\end{center}
\end{inlinefigure}

\newpage
\begin{inlinefigure}
\begin{center}
\resizebox{\textwidth}{!}{\includegraphics{f2.ps}}
\figcaption{ \figtwocap }
\end{center}
\end{inlinefigure}

\newpage
\begin{inlinefigure}
\begin{center}
\resizebox{\textwidth}{!}{\includegraphics{f3.ps}}
\figcaption{ \figthreecap }
\end{center}
\end{inlinefigure}

\fi

\end{document}